\documentclass[%
 reprint,
 amsmath,amssymb,
 aps,
prb,
]{revtex4-2}

\usepackage{microtype}
\usepackage{multirow}
\usepackage{amsmath}
\usepackage{bm}
\usepackage{adjustbox} 
\usepackage{graphicx}
\usepackage{dcolumn}
\usepackage{bm}
\usepackage{siunitx}

\usepackage{color, hyperref}
\definecolor{mygrey}{gray}{0.80}
\definecolor{darkblue}{RGB}{8,81,156}
\hypersetup{colorlinks,breaklinks,
            linkcolor=darkblue,urlcolor=darkblue,
            anchorcolor=darkblue,citecolor=darkblue}

\definecolor{super-dark-green}{RGB}{0,69,41}
\definecolor{super-dark-purple}{RGB}{63,0,125}
\definecolor{super-dark-blue}{RGB}{8,48,107}
\definecolor{super-dark-red}{RGB}{165,0,38}

\setcitestyle{super}
\begin{document}


\title{Secondary finite-size effects and multi-barrier free energy landscapes in molecular simulations of hindered ion transport}

\author{Omar Khalifa}
\author{Brian A. Shoemaker}%
\author{Amir Haji-Akbari}
 \email{amir.hajiakbaribalou@yale.edu}
\affiliation{
 Department of Chemical and Environmental Engineering, 
 \\Yale University, New Haven, CT, 06511, USA
}

\date{\today}

\begin{abstract}
\noindent
Ion transport through nanoscale channels and pores is pivotal to numerous natural processes and industrial applications. Experimental investigation of the kinetics and mechanisms of such processes is, however, hampered by the limited spatiotemporal resolution of existing experimental techniques. While molecular simulations have become indispensable for unraveling the underlying principles of nanoscale transport, they also suffer from some important technical limitations. In our previous works, we identified strong polarization-induced finite-size effects in molecular dynamics simulations of hindered ion transport, caused by spurious long-range interactions between the traversing ion and the periodic replicates of other ions. To rectify these artifacts, we introduced the Ideal Conductor/Dielectric Model (\textsc{Icdm}), which treats the system as a combination of conductors and dielectrics, and constructs an analytical correction to the translocation free energy profile. Here, we investigate some limitations of this model. Firstly, we propose a generalized approach based on Markov State models that is capable of estimating translocation timescales in the thermodynamic limit  for free energy profiles with multiple comparable barriers. Second, we identify a new category of polarization-induced finite-size effects, which significantly alter the spatial distribution of non-traversing ions in smaller systems. These secondary effects cannot be corrected by the ICDM model and must be avoided by selecting sufficiently large system sizes. Additionally, we  demonstrate through multiple case studies that finite-size artifacts can reverse expected trends in ion transport kinetics. Our findings underscore the necessity for careful selection of system sizes and the judicious application of the \textsc{Icdm} model to rectify residual finite-size artifacts.
\end{abstract}

\maketitle


\section{Introduction}
\label{section:intro}
\noindent
Ion transport through nanoscale channels and pores plays a pivotal role in both natural processes and engineered systems. In biology, ion channels\cite{AckermanNewEngJMed1997, MacKinnonFEBSLett2003} constitute a crucial class of  proteins that get embedded in cellular and organellar membranes. Thanks to their superior selectivity that directly stems from their unique structures, such channels regulate the transport of various ions to fulfill important physiological functions. Tuning ion transport is also fundamental to numerous technological applications, such as biosensing,\cite{CornellNature1997, AhmedAnalChem2024} metal recovery,\cite{GebreslassieWaterRes2024} water desalination,\cite{WerberNatRevMater2016} and memristors.\cite{XuACSNano2024}  In all such applications, nano-channels and nanoporous membranes are meticulously engineered to achieve  the specific ion-solvent and ion-ion selectivity needed for their function.

Our ability to understand how biological channels function and to tune the selectivity of synthetic membranes relies on obtaining a molecular-level understanding of how the chemistry and geometry of a pore influence its selectivity. Despite major advancements in the resolution and precision of experimental techniques, they mostly lack the spatiotemporal resolution needed for probing individual ion transport events. Consequently, molecular simulations have emerged as vital complementary tools, offering detailed insights into the physics of nanoscale transport and enabling a systematic exploration of extensive design spaces to discern the principles of structure-selectivity relationships.
 Over recent decades, molecular dynamics (\textsc{Md}) simulations have been pivotal in investigating ion and water transport through various materials including two-dimensional (2D) materials,\cite{CohenTanugiNanoLett2012, HeiranianNatComm2016, ShoemakerACSNano2024} carbon nanotubes,\cite{MaNatureNanotech2015, LynchChemRev2020} and metal-organic frameworks (\textsc{Mof}s).\cite{ZhangSciAdv2018} These simulations have not only provide mechanistic explanations of experimental observations but also predicted new phenomena not previously observed in experiments.

While \textsc{Md} simulations have provided pivotal insights into the fundamental physics of nanoscale transport, they  are inherently constrained by several technical limitations. One important limitation stems from the fact that typical systems considered in \textsc{Md} are often comprised of $10^4-10^6$ particles, which is orders of magnitude smaller than real-world systems. Such small system sizes result in systematic deviations from the thermodynamic limit, known as \emph{finite-size effects}, impacting the reliability of estimates of a wide variety of properties, such as diffusivities,\cite{JamaliJCTC2018, CelebiMolSimul2021} thermal \cite{ChantrenneJHeatTransfer2004, GrasselliJCP2022} and structural properties,\cite{SalacusePhysRevE1996} nucleation rates,\cite{WedekindJCP2006, HussainJCP2021, HussainJCP2022} and glass dynamics.\cite{HorbachPRE1996}  The principal cause of finite-size effects is the use of periodic boundary conditions (\textsc{Pbc}s), which are widely employed in molecular simulations to avoid interfacial artifacts along dimensions with no confinement. Despite providing a practical solution,  \textsc{Pbc}s are inherently unphysical since they entail replicating the system periodically along one or more dimensions.

In our prior work,\cite{ShoemakerJCTC2022} we discovered a novel category of finite-size effects in \textsc{Md} simulations of hindered ion transport through nanoporous membranes that separate two reservoirs. When an ion traverses the pore, it  polarizes the other ions in the reservoirs, leading to finite-size artifacts caused by spurious interactions between the traversing ion and the periodic replicates of the other ions. These artifacts can be substantial, and can alter the transport timescales by several orders of magnitudes. To mitigate these artifacts, we introduced the ideal conductor model,\cite{ShoemakerJCTC2022} which treats the electrolytic solution as an ideal conductor and the membrane as a low-dielectric medium, calculating analytically the correction to the translocation free energy. Subsequently, we extended this model to accommodate dielectric heterogeneities. The generalized model, called the ideal conductor/dielectric model (\textsc{Icdm}),\cite{ShoemakerJCP2024} outperforms the original ideal conductor model. Moreover, it can successfully handle secondary ion transport events, which was not possible with the original model.

In this work, we examine several limitations of the \textsc{Icdm} model. First, we revisit the question of estimating translocation timescales in the thermodynamic limit by comparing the original and \textsc{Icdm}-corrected free energy profiles, which we had previously handled using the Arrhenius relationship. We demonstrate the inefficacy of the Arrhenius approach in situations where there are multiple comparable free energy barriers in the original or corrected free energy profiles, and develop an alternative approach based on analyzing the spectral properties of a  Markov state model (\textsc{Msm}). Secondly, we discover a new class of polarization-induced finite-size effects in ion transport simulations, which cannot be fully resolved by the \textsc{Icdm} model. These artifacts, which we call \emph{secondary finite-size effects}, emerge from subtle-- yet significant-- changes in the spatial distribution of trailing ions in systems that are too small, and result in a significant alteration of the fundamental physics of ion translocation. Unlike the primary artifacts identified in our earlier works that persist in all finite simulations, secondary artifacts vanish in sufficiently large systems, allowing us to propose a heuristic to circumvent them. Finally, we illustrate through multiple examples the varying scaling of finite-size effects with system size, which complicates the common practice of rank ordering the behaviors of systems with differing chemistries but identical cross-sectional surface areas.

This paper is organized as follows. A brief overview of the \textsc{Icdm} model is provided in Section~\ref{section:ICDM}. Section~\ref{section:methods} details the technical aspects of \textsc{Md} simulations, rate calculations, and the numerical implementation of the \textsc{Icdm} model. Our findings are presented in Section~\ref{section:results}, followed by concluding remarks in Section~\ref{section:conclusions}.

\section{Overview of the ICDM Model}
\label{section:ICDM}

\noindent
In the \textsc{Icdm} model,\cite{ShoemakerJCP2024}  the simulation box is partitioned into multiple dielectric domains separated by flat, infinitely large interfaces assumed to be perpendicular to the $z$ axis. If a domain is occupied by an electrolyte at a sufficiently high concentration, it is treated as an ideal conductor, mathematically represented as possessing a dielectric constant of $\epsilon_r=\infty$. Otherwise, its dielectric constant is estimated numerically.  For all charges located within non-conducting domains, including the traversing ion, their image charges are enumerated using the method of images.  This allows for the estimation of $E^{\text{ex}}(z)$, the $z$ component of the excess electric field exerted on the traversing ion by the periodic replicates of all other real and image charges in the system.  The correction to the translocation free energy is then estimated as follows:
\begin{eqnarray}
\Delta\mathcal{F}_{\text{corr}}(z) = -\int_{z_0}^{z} q_tE^{\text{ex}}_z\left(\overline{z}\right)\,d\overline{z},\label{eq:corr-general}
\end{eqnarray}
where $q_t$ is the charge of the traversing ion. The excess electric field is estimated in a modular manner as follows. For localized charges within the pore, a point charge correction is applied, which, for the traversing ion, is given by:
\begin{eqnarray}
{\footnotesize
E_{\text{pt},z}^{\text{ex}}(z_t) = \sum_{i=1}^{+\infty}\sum_{\begin{array}{l}\mathbf{m}\in\mathbb{Z}^2\\ \mathbf{m}\neq\pmb0\end{array}}
\frac{q_t(z_t-z_i)}{4\pi\epsilon_0\epsilon_p
\left|(m_xL_x,m_yL_y,z_t-z_i)\right|^3
}}\notag \\
\label{eq:Ez-pt-self}
\end{eqnarray}
Here, $\epsilon_p$ is the dielectric constant of the domain containing the pore, $L_x$ and $L_y$ are the dimensions of the periodic box along the $x$ and $y$ directions,  $|\cdot|$ represents the Euclidean norm, and $q_i$'s and $z_i$'s correspond to the magnitudes and positions of all of $q_t$'s image charges. For other non-traversing charges within the pore, the point-charge correction will also include the contributions from their own periodic replicates:
\begin{eqnarray}
E_{\text{pt},z}^{\text{ex}}(z_t|q_j,z_j) &=& \frac{1}{4\pi\epsilon_0\epsilon_p}\sum_{
\begin{array}{c}
\mathbf{m}\in\mathbb{Z}^2\\
\mathbf{m}\neq\pmb0
\end{array}
}\Bigg[
\notag\\
&&
\frac{q_i(z_t-z_j)}{\left|(m_xL_x,m_yL_y,z_t-z_j)\right|^3}\notag\\
&& +\sum_{i=1}^{\infty}\frac{q_{j,i}(z_t-z_{j,i})}{\left|(m_xL_x,m_yL_y,z_t-z_{j,i})\right|^3}\Bigg]\label{eq:Ez-pt-other}
\end{eqnarray}
Here, $q_j$ and $z_j$ correspond to the magnitude and the position of the $j$-th charge within the pore, and $q_{j,i}$'s and $z_{j,i}$'s are the magnitudes and positions of its image charges, respectively.

As for the feed and filtrate domains, slab corrections are applied:
\begin{eqnarray}
E_{\text{slab},z}^{\text{ex}}(z_t|\sigma_f,z_f) &=& \frac{\sigma_f}{4\pi\epsilon_0\epsilon_p}\Bigg[
2\pi-\notag\\
&&\int_{-\frac{L_x}2}^{\frac{L_x}2}\int_{-\frac{L_y}2}^{\frac{L_y}2}\frac{(z_t-z_f)\,dx\,dy}{\left|(x,y,z-z_p)\right|^3}
\Bigg], \label{eq:Ez-slab}
\end{eqnarray}
where $\sigma_f$ and $z_f$ correspond to the surface charge density and the location of the slab, respectively.  In cases where the feed or filtrate is conducting, slab corrections account only for excess charges accumulated at their surfaces, excluding bound interfacial charges from image estimates. A dielectric feed or filtrate is, however, partitioned into smaller bins, with the surface charge density within each bin estimated from \textsc{Md} simulations. 

\begin{table*}[hbt!]
    \centering
    \caption{\label{tab:sys-summary}Summary of system sizes considered in this work. $L_x$ and $L_y$ correspond to the dimensions of the simulation box along the $x$ and $y$ directions, while $N_{w,\text{feed}}$ and $N_{w,\text{filtrate}}$ are the number of water molecules in the feed and filtrate, respectively. $N_{i}$ is the number of ion pairs in the system.}
    \begin{tabular}{lcccccc}
        \hline\hline
        System & $L_x\,(\text{nm})$ & $L_y\,(\text{nm})$ & $\text{Area}\,(\text{nm}^2)$ & {$N_{w,\text{feed}}$} & {$N_{w,\text{filtrate}}$} & {$N_{i}$}\\
        \hline
        \multirow{3}{*}{$2\delta$ single-pore systems} & ~$7.614$~ & ~$6.806$~ & ~~$51.822$~~ & ~6,473 & ~4,380 & 180 \\
                                 & $10.070$~ & $10.210$~ & ~$102.808$~~ & 13,275~ & ~8,982~ & 371 \\
                                 & $14.245$~ & $14.464$~ & ~$206.035$~~ & 25,737~ & 17,416~ & 716 \\
        \hline
        \multirow{4}{*}{\shortstack{$1\delta$ single-pore systems\\ (all transitions)}} & ~$3.684$~ & ~$3.403$~ & ~~$12.537$~~ & ~1,616~ & ~1,095~ & ~45 \\
                                 & ~$5.158$~ & ~$5.105$~ & ~~$26.329$~~ & ~3,393~ & ~2,300~ & ~95 \\
                                 & ~$6.386$~ & ~$5.105$~ & ~~$32.598$~~ & ~4,220~ & ~2,857~ & 118 \\
                                 & $10.070$~ & $10.210$~ & ~$102.808$~~ & 13,275~ & ~8,982~ & 371 \\
        \hline
        \multirow{2}{*}{$1\delta$ double-pore systems} & ~$6.386$~ & ~$5.105$~ & ~~$32.598$~~ & ~4,220~ & ~2,857~ & 118 \\
                                 & $10.070$~ & $10.210$~ & ~$102.808$~~ & 13,275~ & ~8,982~ & 371 \\
        \hline
    \end{tabular}
\end{table*}

For any given membrane system, the $E_z^{\text{ex}}(z)$ of Eq.~\eqref{eq:corr-general} can be readily constructed by combining a sufficient number of point and slab corrections. For instance,  for the translocation of a first ion from a conducting feed to an ion-free filtrate, the correction is given by:
\begin{eqnarray}
E_z^{\text{ex}}(z_t) &=& E_{\text{pt},z}^{\text{ex}}(z_t) + \sum_{j=1}^N E_{\text{pt},z}^{\text{ex}}(z_t|q_j,z_j)\notag\\
&& + \sum_{k=1}^2E_{\text{slab},z}^{\text{ex}}\left(
z_t|\frac{q_e}{2L_xL_y},z_{c,k}
\right)
\end{eqnarray} 
Here, $q_e$ is the excess charge within the feed, and $z_{c,1}$ and $z_{c,2}$ are the locations of its two surfaces, the piston and the membrane surface, respectively.  $N$ is the number of other charges present within the pore. When it comes to non-traversing point charges within the pore, it is more practical to discretize the membrane and compute the non-traversing charge density profile as a function of the traversing ion's position. Consequently, in practice, $N$ will correspond to the number of spatial bins within the pore, and $q_j$'s will be the average charge content of the $j$-th bin.

It is necessary to emphasize that the \textsc{Icdm} model is a post-processing tool that estimates $\Delta\mathcal{F}_{\text{corr}}(z)$ by post-processing \textsc{Md} trajectories to obtain non-traversing ionic densities, and other relevant observables. $\mathcal{F}_f(z)$, the \emph{uncorrected} translocation free energy profile is determined separately from the aforementioned trajectories. $\mathcal{F}_{\infty}(z)$, the free energy profile in the thermodynamic limit is then estimated by subtracting $\Delta\mathcal{F}_{\text{corr}}(z)$  from $\mathcal{F}_f(z)$.

It is also important to note that the \textsc{Icdm} model does not explicitly account for the kinetics of translocation. Indeed, the sampling method used to compute $\mathcal{F}_f(z)$ can be purely equilibrium-based, without necessarily providing kinetic information, thereby allowing for the use of bias-based methods such as umbrella sampling\cite{TorrieJComputPhys1977} or metadynamics.\cite{LaioPNAS2002} In our prior works,\cite{ShoemakerJCTC2022, ShoemakerJCP2024} we employed forward-flux sampling\cite{AllenJCP2006, HussainJChemPhys2020} (\textsc{Ffs}), a path sampling technique that allows for simultaneous calculation of passage times and translocation free energy profiles. In order to estimate the passage times in the thermodynamic limit, we introduced an Arrhenius-type framework predicated on several approximations, most notably the assumption that the translocation process is predominantly characterized by the crossing of a single dominant free energy barrier. Section~\ref{section:Markov} examines this approximation and proposes a more generalized \textsc{Msm}-based alternative for conducting such an extrapolation.

A key assumption of the \textsc{Icdm} model is that the fundamental physics of the ion transport process is not significantly perturbed in the finite system. In other words, it presumes that the relevant observables of the finite system are qualitatively similar to those in an infinite system. Our work, however, demonstrates that this is not always the case and that, in very small systems, the transport physics can change substantially. As will be discussed in Section~\ref{section:secondary}, these secondary finite-size effects can affect the accuracy of the correction, leading to a systematic misestimation of corrected translocation barriers and passage times. This makes it essential to devise strategies for avoiding such artifacts, as outlined in  Section \ref{section:heuristic}.

\begin{figure*}[t]
	\centering
	\includegraphics[width=1\textwidth]{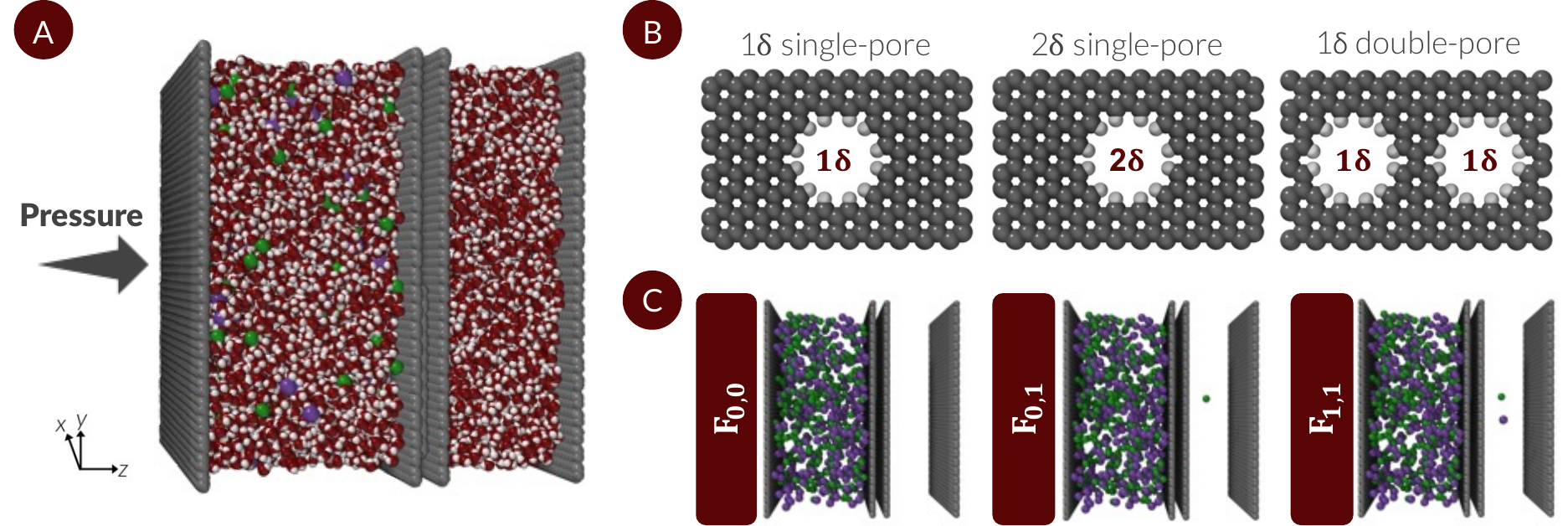 }
	\caption{\label{fig:setup}
	(A) A representative snapshot of the simulation box, with a three-layer graphitic membrane separating a feed reservoir comprised of sodium (purple) and chloride (green) ions and water molecules, from a filtrate reservoir containing water molecules only.  Both reservoirs are surrounded by single-layer graphene pistons. (B) The membranes investigated here include two single-pore membranes  with 1$\delta$ or 2$\delta$ partial charges on the hydrogen atoms (white), and a double-pore $1\delta$ membrane.  (C) Schematic representation of the investigated starting and target basins, indexed as $F_{p,q}$, representing states wherein $p$ sodium and $q$ chloride ions are present in the filtrate. Water molecules are removed for clarity.}
\end{figure*}

\section{Methods}
\label{section:methods}

\subsection{System setup}
\noindent
The methodologies employed in this work to construct the systems and conduct the simulations closely follow  the approaches detailed in  our earlier works.\cite{MalmirMatter2020,ShoemakerJCTC2022,ShoemakerJCP2024,ShoemakerACSNano2024}  All systems considered in this work consist of passivated sub-nanometer pores etched within graphitic sheets. First, a pristine armchair graphene sheet is generated using the \texttt{Nanotube Builder} extension of \textsc{Vmd}.\cite{HumphreyJMolGraph1996} The sheet is subsequently perforated by removing carbon atoms within a distance 4.3~\AA~from the center of a single  hexagonal ring.  By employing the approach outlined in Ref.~\citenum{ShoemakerJCTC2022}, we assure that only single bonds are trimmed during the perforation process. Carbon atoms with dangling bonds are passivated with hydrogen atoms. To investigate the effect of pore interior polarity, we consider two pore types. In addition to passivating hydrogens with a standard partial charge of $+0.115e$\cite{MalmirMatter2020} (designated as $1\delta$), we consider more polar pores with double the standard charge $2\delta=+0.230e$ (denoted as $2\delta$ pores). In both scenarios, the passivated carbons are assigned an opposite partial charge to maintain charge neutrality.

Each membrane is comprised of three stacked perforated graphene sheets, spaced 3.35~\AA~apart, with the middle layer laterally shifted by 1.418~\AA~to replicate the packing structure of graphite (Fig.~\ref{fig:setup}A). We consider three types of membranes: the \textit{single-pore} and \textit{double-pore} $1\delta$ membranes, and the single-pore $2\delta$ membrane (Fig.~\ref{fig:setup}B).  Each membrane separates two reservoirs: the feed comprised of an aqueous NaCl solution, and the filtrate, which contains only water molecules.  Solution configurations are generated using \textsc{Packmol}.\cite{MartinezJComputChen2009} Subsequently, pristine graphene sheets are added to both sides of the feed-membrane-filtrate structure to act as pistons (see Figure~\ref{fig:setup}). To ensure adequate sampling of the configurational space, 100 to 200 independent initial configurations are prepared by using different random number generator seeds for the insertion of water molecules and Na$^+$ and Cl$^-$ ions. Table~\ref{tab:sys-summary} summarizes the dimensions of the simulation box, as well as the number of water molecules and ion pairs present within each reservoir for all systems considered in this work.

\subsection{Molecular dynamics simulations}
\label{section:methods:MD}

\noindent
All atoms are represented as Lennard-Jones (\textsc{Lj}) particles with partial charges. The \textsc{Tip3p}\cite{PriceJChemPhys2004} and Joung-Cheatham (\textsc{Jc})\cite{JoungJPhysChemB2008} force-fields are used for the water molecules and sodium/chloride ions, respectively. The \textsc{Lj} parameters for carbon and hydrogen atoms in the membrane and pistons are obtained from Beu\cite{BeuJChemPhys2010} and M\"{u}ller-Plathe.\cite{MullerPlathMacromolecules1996} All \textsc{Md} simulations are conducted using the Large-scale Atomic/Molecular Massively Parallel Simulator (\textsc{Lammps}),\cite{ThompsonComputPhysCommun2022} where the equations of motions are propagated using the velocity-Verlet algorithm.\cite{SwopeJCP1982} Membrane atoms are fixed throughout the simulation. The \textsc{Shake} algorithm\cite{AndersenJComputPhys1993} is used to maintain the rigidity of the water molecules. Periodic boundary conditions are applied along the $x$ and $y$ directions only in order to avoid unphysical interactions between the system and its periodic images along the $z$ direction. As such, the long-range electrostatic interactions are calculated using the slab particle-particle particle-mesh (\textsc{Pppm})\cite{BostickBiophysJ2003} method. The Nosé-Hoover thermostat\cite{NoseMolPhys1984, HooverPhysRevA1985} with a damping time constant of 0.1~ps is utilized for controling the temperature at 300 K. We equilibrate the system for 0.1~ns at the \textsc{Nvt} ensemble using a time step of 0.5~fs while the pistons are fixed. The pistons are then allowed to move along the $z$ direction using the \textsc{Lammps} \texttt{fix aveforce} command with an effective pressure gradient of 194 atm. We equilibrate the system with moving pistons for an additional 2~ns before starting rate calculations.

\begin{table*}[hbt!]
    \centering
    \caption{\label{tab:icdm-summary}Summary of \textsc{Icdm} parameters for different systems. $\epsilon_p$ and $\epsilon_f$ correspond to the dielectric constants of the membrane and the filtrate, respectively, both calculated using Eq.~(\ref{eq:dielec}). The \textsc{Icdm} correction is applied starting at $z_0$.  $z_{c,1}$ and $z_{c,2}$ correspond to the locations of conductor's surfaces at the membrane and the feed piston, respectively whi;e $z_M$ is the locus of the filtrate-piston interface. $n_m$ and $n_f$ are the number of spatial bins in the membrane and filtrate regions, respectively. The parameters $z_M$, $n_m$, and $n_f$ are only relevant when there are non-transiting ions inside the pore or in the filtrate.}
    \begin{tabular}{c|c|cccccccccc}
        \hline\hline
        System type & Transition & $A\,(\text{nm}^2)$ & $\epsilon_p$ & $\epsilon_f$ & $z_0 \ (\text{nm})$ & $z_{c,1} \ (\text{nm})$ & $z_{c,2}$ \ (\text{nm}) & $z_m \ (\text{nm})$ & $z_M \ (\text{nm})$ & $n_m$ & $n_f$\\
        \hline
        \multirow{3}{*}{$2\delta$ single-pore} & $F_{0,0}\rightarrow F_{0,1}$ & $51.822$ & $1.086$ & $70.709$ & $0$ & $-0.582 \pm 0.038$ & $-4.261$ & $8.7$ & $3.62$ & $8$ & $6$\\
                                & $F_{0,0}\rightarrow F_{0,1}$ & $102.808$ & $1.045$ & $63.958$ & $0$ & $-0.611 \pm 0.042$ & $-4.400$ & $8.7$ & $3.72$ & $8$ & $6$\\
                                & $F_{0,0}\rightarrow F_{0,1}$ & $206.035$ & $1.023$ & $68.851$ & $0$ & $-0.628\pm 0.078$ & $-4.291$ & $8.7$ & $3.63$ & $8$ & $6$\\
        \hline
        \multirow{5}{*}{\shortstack{$1\delta$ single-pore}} & $F_{0,1}\rightarrow F_{1,1}$ & $12.537$ & $1.407$ & $71.942$ & $0$ & $-0.460 \pm 0.005$ & $-4.338$ & $8.7$ & $3.83$ & $8$ & $6$\\
                                 & $F_{0,1}\rightarrow F_{1,1}$ & $26.329$ & $1.155$ & $72.798$ & $0$ & $-0.432\pm0.011$ & $-4.423$ & $8.7$ & $3.55$ & $8$ & $6$\\
                                 & $F_{0,0}\rightarrow F_{0,1}$ & $32.598$ & $1.145$ & $71.380$ & $0$ & $-0.483\pm0.022$ & $-4.225$ & $8.7$ & $\dots$ & $\dots$ & $\dots$\\
                                 & $F_{0,0}\rightarrow F_{0,1}$ & $102.808$ & $1.050$ & $71.380$ & $0$ & $-0.452\pm0.021$ & $-4.325$ & $8.7$ & $\dots$ & $\dots$ & $\dots$\\
                                 & $F_{0,1}\rightarrow F_{1,1}$ & $102.808$ & $1.046$ & $71.360$ & $0$ & $-0.185\pm0.139$ & $-4.415$ & $8.7$ & $3.48$ & $8$ & $6$\\
        \hline
        \multirow{2}{*}{$1\delta$ double-pore} & $F_{0,0}\rightarrow F_{0,1}$ & $32.598$ & $1.272$ & $71.505$ & $0$ & $-0.470\pm0.010$ & $-4.367$ & $8.7$ & $\dots$ & $\dots$ & $\dots$\\
                                 & $F_{0,0}\rightarrow F_{0,1}$ & $102.808$ & $1.095$ & $69.956$ & $0$ & $-0.460\pm0.034$ & $-4.414$ & $8.7$ & $3.74$ & $8$ & $6$\\
        \hline
    \end{tabular}
\end{table*}

\subsection{Path sampling simulations and rate calculations}

\noindent
Ion translocation through the nanopores considered in this study is a rare event that  occurs over timescales that cannot be efficiently captured using conventional non-equilibrium \textsc{Md} (\textsc{Nemd}) simulations. To overcome this challenge, we employ the jumpy forward flux sampling\cite{HajiAkbariJChemPhys2018} (\textsc{jFfs}) algorithm to access such extended timescales, with implementation details outlined in our earlier works.\cite{MalmirMatter2020, ShoemakerJCTC2022}

Ion translocation processes are represented as $F_{p,q} \rightarrow F_{p\pm1,q\pm1}$, where the state $F_{p,q}$ corresponds to configurations with $p$ sodium ions and $q$ chloride ions in the filtrate (Fig.~\ref{fig:setup}C). In this study, we focus on two specific transitions: $F_{0,0} \rightarrow F_{0,1}$ and $F_{0,1} \rightarrow F_{1,1}$, which correspond to the translocation of the first chloride ion and the subsequent transport of the first sodium ion, respectively. We use the curved directed distance of the transitioning ion from the pore mouth as the \textsc{jFfs} order parameter, with precise definitions provided in our earlier work.\cite{MalmirMatter2020}

Equilibrated configurations obtained from the procedure described in Section~\ref{section:methods:MD} serve as starting points for exploring the $F_{0,0}$ basin. For the $F_{0,1}\rightarrow F_{1,1}$ transition, initial configurations are randomly selected from the crossings at the final \textsc{Ffs} milestone for the $F_{0,0}\rightarrow F_{0,1}$ transition. The starting basins are sampled for a minimum combined duration of $\approx1\,\mu$s, and individual \textsc{jFfs} iterations are terminated after recording a minimum of 2,000 crossings.   The uncorrected free energy profiles are computed using the forward-flux sampling mean first passage time (\textsc{Ffs-Mfpt}) method,\cite{ThaparJChemPhys2015} with mean first passage times calculated using the jump-corrected expression derived in our earlier work.\cite{ShoemakerJCTC2022}

\subsection{Application of the  ICDM model}

\noindent
The \textsc{Icdm} model outlined in Section~\ref{section:ICDM} is used for estimating passage times and translocation barriers at the thermodynamic limit, using the implementation details outlined in our earlier work\cite{ShoemakerJCP2024} unless stated otherwise. The feed compartment is always treated as an ideal conductor, while the filtrate and the membrane are represented as dielectric media. The dielectric constants of these regions are calculated using fluctuations in dipole moments according to the following expression:\cite{Neumann1MolPhys983}
\begin{equation}
    \epsilon = 1 + \frac{\langle |\textbf{M}|^{2}\rangle - \left|\langle\textbf{M}\rangle\right| ^{2}}{3\epsilon_{0}kT\langle V\rangle}\label{eq:dielec}
\end{equation}
where $\epsilon_0$ is the vacuum permittivity, $k$ is the Boltzmann constant, and $\langle V \rangle$ is the volume of the region of interest.  It is essential to note that in calculating the dipole moment of water molecules, a molecule is considered to be within a specific domain if its oxygen atom resides within that domain. All corrections are executed using $n_{im}=2000$ image charges and $n_{prd}=200$ periodic replicates unless otherwise specified. The remaining model parameters are given in Table~\ref{tab:icdm-summary}.

\section{Results and Discussion}
\label{section:results}

\subsection{A Markov state approach for estimation of passage times in the thermodynamic limit}
\label{section:Markov}

\noindent
In our prior works,\cite{ShoemakerJCTC2022, ShoemakerJCP2024} we estimated ionic passage times in the thermodynamic limit by assuming the presence of a single dominant free energy barrier in both uncorrected and corrected profiles. The passage time in the thermodynamic limit, $\tau_\infty$, was then estimated by:\cite{KramersPhysica1940}
\begin{eqnarray}
\frac{\tau_\infty}{\tau_f} =e^{\beta\left[\Delta\mathcal{F}_\infty-\Delta\mathcal{F}_f
\right]}.\label{eq:Arrhenius}
\end{eqnarray}
However, ion transport through nanopores  is often complex, and could  involve free energy profiles with multiple peaks and valleys. Here, we propose a more generalized approach based on Markov state models\cite{HusicJACS2018} to handle such scenarios, particularly when the free energy profile features multiple comparable barriers. 

\begin{table}
\centering
\caption{\label{tab:MS-old}Comparison between the Arrhenius and \textsc{Msm} estimates of $\tau_\infty$ of the $F_{0,0}\rightarrow F_{0,1}$  transition in the $1\delta$ system. The analysis utilized data reported in our earlier work.\cite{ShoemakerJCP2024}}
\begin{tabular}{c|c|ll}
\hline\hline
$A\,(\text{nm}^2)$ & $\tau_f\,(\text{ns})$ & $\tau_{\infty,\text{Arrhenius}}\,(\text{ns})$ & $\tau_{\infty,\textsc{Msm}}\,(\text{ns})$ \\
\hline
~$12.53$ & $(8.29\pm0.53)\times10^7$ & ~~~~~$99\pm24$ & ~~$61\pm4$ \\
~$26.33$ & $(1.36\pm0.17)\times10^4$ & ~~~~$112\pm22$ & ~~$78\pm10$\\
~$32.60$ & $(3.38\pm0.17)\times10^3$ & ~~~~$141\pm27$ & ~~$62\pm4$\\
~$51.82$ & $(4.69\pm0.58)\times10^2$ & ~~~~~$96\pm17$ & ~~$50\pm6$\\
$102.81$ & $(1.12\pm0.16)\times10^2$ & ~~~~~$49\pm22$ & ~~$51\pm10$\\
\hline
\end{tabular}
\end{table}

To justify the need for our \textsc{Msm} model,  it is necessary to highlight the inherent challenges in computing mean first passage times (\textsc{Mfpt}s) along a reaction coordinate, as the well-established theoretical framework for this task is of limited practical utility.  Assuming that the underlying system evolves diffusively along a one-dimensional reaction coordinate, $\lambda$, the conditional probability density $p(\lambda,t|\lambda_0,0)$ will be governed by the Smoluchowski equation:\cite{DominguesJCTC2024}
\begin{eqnarray}
\frac{\partial p}{\partial t} &=& \frac{\partial}{\partial\lambda}\left[
D(\lambda)e^{-\beta\mathcal{F}(\lambda)}\frac{\partial}{\partial\lambda}\left[e^{\beta\mathcal{F}(\lambda)}p\right]
\right],\label{eq:Smoluchowski}
\end{eqnarray}
where $\mathcal{F}(\lambda)$ and $D(\lambda)$ represent the free energy and diffusivity profiles, respectively. The \textsc{Mfpt} between two successive basins, $\lambda_A$ and $\lambda_B$, can be estimated from:\cite{HinczewskiJCP2010}
\begin{eqnarray}
\tau_{A\rightarrow B} = \int_{\lambda_A}^{\lambda_B}\int_{\lambda_A}^{\xi} \frac{e^{\beta\left[\mathcal{F}(\xi)-\mathcal{F}(\zeta)\right]}}{D(\xi)}\,d\zeta\,d\xi.
\label{eq:MFPT-general}
\end{eqnarray}
Unfortunately, Eq.~\eqref{eq:MFPT-general} is of limited practical utility, since even though it is fairly straightforward to obtain $\mathcal{F}(\lambda)$ from \textsc{Md} trajectories,  the underlying diffusivity profile $D(\lambda)$ is often not known. To render Eq.~\eqref{eq:MFPT-general} more tractable, one can simplify it by  assuming that $D(\lambda)$ is constant, and estimating $\tau_{A\rightarrow B}$ from $\mathcal{F}(\lambda)$. If the free energy barrier separating the two basins is sufficiently large, Eq.~\eqref{eq:MFPT-general} can be further simplified to yield:
\begin{eqnarray}
\tau_{A\rightarrow B} \approx Ae^{\beta\left[\mathcal{F}(\lambda_l)-\mathcal{F}(\lambda_A)\right]} .
\label{eq:MFPT-simplified}
\end{eqnarray}
Here, $\lambda_l$ is the locus of the barrier, while $A$ is a prefactor that depends on the curvatures of the barrier and the starting basin, and the average diffusivity. Notably, Eq.~\eqref{eq:MFPT-simplified} corresponds to the renowned Arrhenius relationship, employed in our previous work. 

Due to the challenges of applying Eq.~\eqref{eq:MFPT-general} to multi-barrier free energy profiles, we propose a simple \textsc{Msm} model with $m$ states corresponding to the local minima of $\mathcal{F}(\lambda)$. The $m$-th state represents the target basin and is treated as an absorbing boundary, as our objective is to estimate the time required to transition from any other state to state $m$. The temporal evolution of the state occupancy vector,  $\mathbf{p}(t)=[p_1(t),\cdots,p_m(t)]$, which we describe as a row vector, can be described by the rate equation:
$$d{\mathbf{p}/dt} = \mathbf{p}\cdot\mathbf{R},$$
with $\mathbf{R}$ a tridiagonal rate matrix with the following entries:
\begin{eqnarray}
R_{ij} &=& \left\{
\begin{array}{ll}
k_{i,i-1} & i=j+1 \\
k_{i+1,i} & i=j-1\\
-k_{i+1,i}-k_{i-1,i} & 1<i=j<m\\
-k_{2,1} & i=j=1\\
0 & \text{otherwise}
\end{array}
\right..
\end{eqnarray}
Here, $k_{i,i\pm1}$ is the rate of transitioning from state $i$ to $i\pm1$. Assuming the validity of the Arrhenius equation for transitions between successive basins, we can estimate $k_{i,i\pm1}$ as:
\begin{eqnarray}
k_{i,i\pm1} &=& e^{\beta\left[\mathcal{F}_i-\mathcal{F}^*_{i\rightarrow i\pm1}\right]},
\label{eq:rate-constant}
\end{eqnarray}
with $\mathcal{F}^*_{i\rightarrow i\pm1}$ denoting the magnitude of the barrier separating the two states. It is important to note that no prefactor is included in Eq.~\eqref{eq:rate-constant} since our aim is to compare the relative dynamics of $\mathcal{F}_f(\lambda)$ and $\mathcal{F}_\infty(\lambda)$. Additionally, $k_{m,m-1}$ is set to zero, reflecting the assumption that the $m$-th state is an absorbing state.

\begin{table*}
\caption{\label{tab:times-2X}Translocation timescales and free energy barriers obtained from \textsc{jFfs}  for the $F_{0,0} \rightarrow F_{0,1}$ transition in the 2$\delta$ system for different system sizes.}
\begin{tabular}{lccccc}
\hline\hline
$A$~($\text{nm}^2$) & $\tau_{f}\,(\text{ns})$ & $\beta \Delta \mathcal{F}_{f}$ &  $\beta \Delta \mathcal{F}_{\infty}$ & $\tau_{\infty,\text{Arrhenius}}\,(\text{ns})$ & $\tau_{\infty,\textsc{Msm}}\,(\text{ns})$ 
\\
\hline
~~$51.82$ & $4290.71\pm231$ & $10.72\pm0.02$ & $2.57\pm0.07$ & $1.24\pm0.15$ & $0.59\pm0.06$\\
~$102.81$&  ~~~$91.93\pm4.28$ & ~$6.53\pm0.01$ & $1.34\pm0.03$ & $0.51\pm0.04$ & $1.41\pm0.18$\\
~$206.04$ & ~~~$14.23\pm0.69$ & ~$4.17\pm0.03$ & $1.53\pm0.11$ & $1.02\pm0.18$ & $1.60\pm0.10$\\
\hline 
\end{tabular}
\end{table*}

  It can be easily shown that $\mathbf{R}$ is a singular matrix, with non-positive eigenvalues. The zero eigenvalue corresponds to the occupancy matrix at $t\rightarrow\infty$, which, due to the absorbing nature of the $m$-th state is given by $[0,0,\cdots,0,1]$.  The remaining negative eigenvalues define the different timescales associated with converging to that distribution. Among these timescales, $\tau_{\text{slow}}$, the largest time constant-- obtained from the smallest nonzero eigenvalue of $\mathbf{R}$ in magnitude-- corresponds to the (relative) ion translocation timescale. Therefore, one can  adjust the $\tau_f$ obtained from direct rate calculations as,
\begin{eqnarray}
\tau_\infty = \tau_f \frac{\tau_{\text{slow}}(\mathbf{R}_\infty)}{\tau_{\text{slow}}(\mathbf{R}_f)}\label{eq:MarkovMatrixScaling}
\end{eqnarray}
To evaluate the performance of this approach, we recalculated $\tau_\infty$ for the $F_{0,0}\rightarrow F_{0,1}$ transition in the $1\delta$ system, using the free energy profiles and passage times reported in our previous work.\cite{ShoemakerJCP2024} Table~\ref{tab:MS-old} presents the estimates obtained from Eqs.~\eqref{eq:Arrhenius} and \eqref{eq:MarkovMatrixScaling}. Clearly, the \textsc{Msm} approach yields more consistent passage times that differ by less than a factor of two for the range of system sizes considered therein. In contrast, the Arrhenius-based estimates exhibit wider variability, with the two system sizes, namely 26.33~nm$^2$ and 32.60~nm$^2$, as notable outliers. Interestingly, these systems both possess uncorrected free energy profiles with two comparable barriers. When both uncorrected and corrected free energy profiles possess a single dominant barrier, the two methods provide statistically indistinguishable estimates. 

\begin{figure*}[t]
	\centering
	\includegraphics[width=0.7\textwidth]{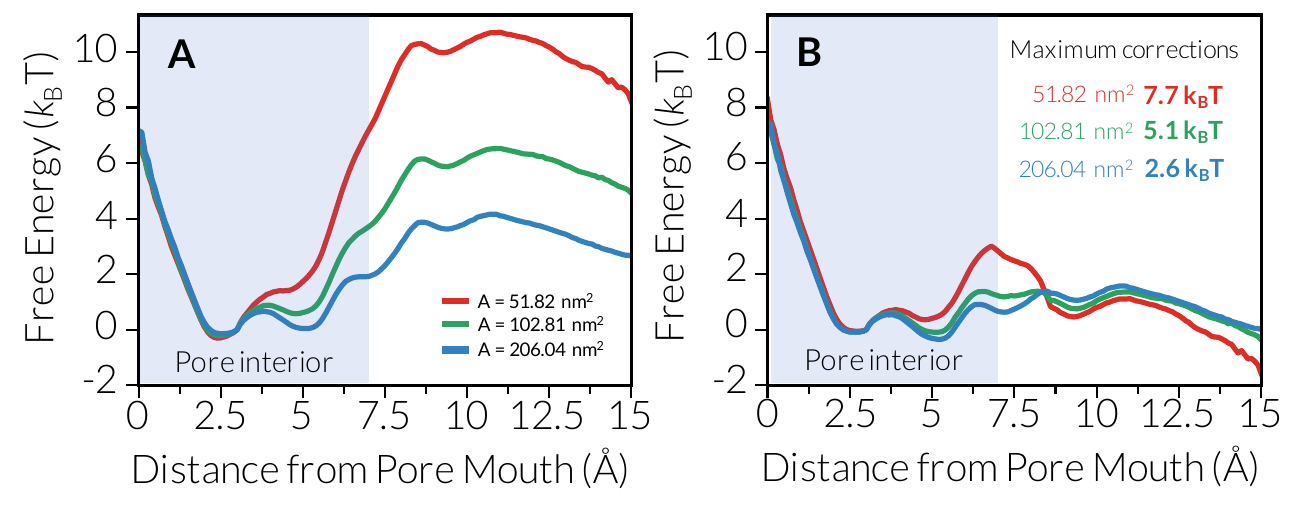}
	\caption{\label{fig:2XCorrections}(A) Uncorrected and (B) \textsc{Icdm}-corrected free energy profiles for the $F_{0,0} \rightarrow F_{0,1}$ transition in the $2\delta$ system. The magnitude of the correction at the top of the uncorrected free energy barrier for each system size is shown in (B). Shaded regions correspond to the pore interior.}
\end{figure*}

\subsection{Secondary finite-size effects}
\label{section:secondary}

\noindent
As discussed in Sections~\ref{section:intro} and \ref{section:ICDM}, secondary finite-size effects originate from systematic alterations in ion transport physics when dealing with systems that are too small. In this section, we illustrate this through two case studies. The first involves the transport  of the leading chloride  through a $2\delta$ pore (Section~\ref{section:2X}), while the second examines sodium  transport through a $1\delta$ pore (Section~\ref{section:sodium}).

\subsubsection{Primary translocation in the presence of trailing ions}
\label{section:2X}

\begin{figure*}[t]
	\centering
	\includegraphics[width=.8\textwidth]{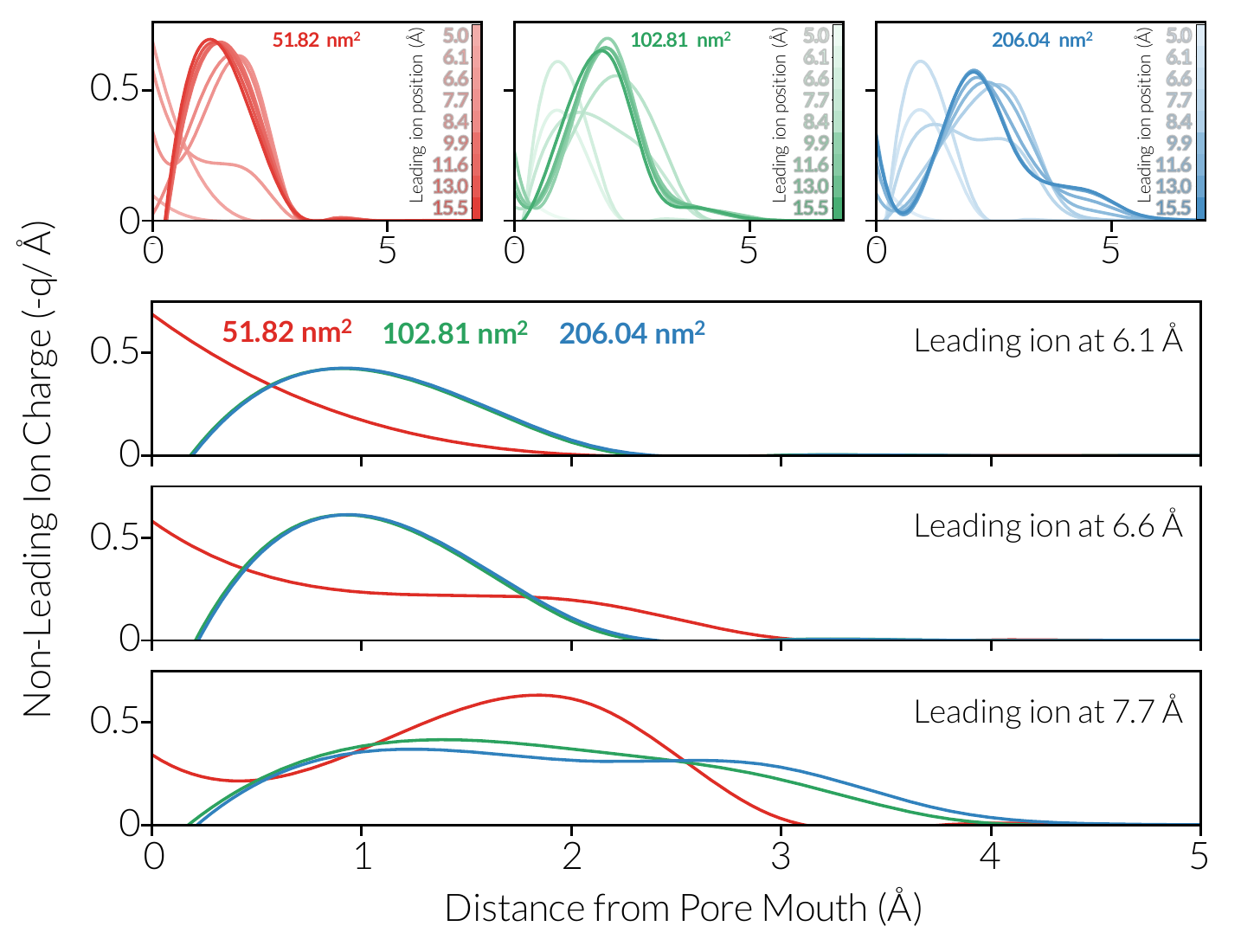}
	\caption{\label{fig:2XTrailingCl} 
	Contributions of the trailing chloride to $\rho_q(z,\lambda)$ for various values of $\lambda$ across different system sizes. The top row displays $\rho_q(z,\lambda)$ for different $\lambda$ values (i.e.,~\textsc{jFfs} milestones) within each system, with $\lambda$ ranging from $5\,\AA$ to $15.5\,\AA$; darker colors represent larger $\lambda$ values. The three bottom rows present $\rho_q(z,\lambda)$ across different system sizes at specific $\lambda$ values: 6.1~\AA, 6.6~\AA, and 7.7~\AA. These $\lambda$'s correspond to points of discrepancy in the free energy profiles of the three system sizes. All curves were generated by binning the pore interiors (1000 bins) and smoothing the computed charge densities using smoothing cubic splines. All $\rho_q(z,\lambda)$ values are normalized by the bin width.}
\end{figure*}

\noindent
We first examine the $F_{0,0}\rightarrow F_{0,1}$ transition through the more polar $2\delta$ pore. What makes this transition  fundamentally different from the one observed in the $1\delta$ pore\cite{ShoemakerJCP2024} is the partitioning of a trailing chloride  into the pore during the translocation of the leading chloride. This distinct behavior can be attributed to the stronger electrostatic attractions between the stronger dipoles within the pore interior and the counter-ions. Consequently, the likelihood of a secondary--  and occasionally a tertiary-- chloride ion entering the pore increases, as depicted in Fig.~\ref{fig:2XTrailingCl}.

As mentioned in Section~\ref{section:ICDM}, the  \textsc{Icdm} model is, in principle, capable of accounting for the finite-size effects induced by non-transiting ions within the membrane.  Figure~\ref{fig:2XCorrections} illustrates the uncorrected (Fig.~\ref{fig:2XCorrections}A) and \textsc{Icdm}-corrected (Fig.~\ref{fig:2XCorrections}B) translocation free energy profiles across different system sizes. Consistent with the $1\delta$ system, the uncorrected translocation barrier, $\Delta\mathcal{F}_f$, decreases with increasing system size. However, the corrected free energy profile of the smallest system (51.82 nm$^2$) diverges substantially  from those of the two larger systems, which exhibit remarkable consistency with each other. Furthermore, visual inspection of the uncorrected profiles reveals a subtle qualitative difference, namely the  emergence of a shoulder at approximately $6.2$~\AA~for the 102.81 nm$^2$ system, evolving into a stationary inflection point in the 206.04 nm$^2$ system. This feature is notably absent in the smallest system, suggesting that the fundamental physics of ion transport might qualitatively differ in the smallest system.

To understand the origin of this discrepancy, we examine the evolution of the spatial distribution of the trailing chloride as the leading chloride traverses the pore. Figure~\ref{fig:2XTrailingCl} illustrates $\rho_q(z,\lambda)$, representing the contribution of trailing ions to the charge density within the pore interior when the leading ion is positioned at a curved directed distance, $\lambda$, from the pore mouth. Examining the profiles for 6.1$\,$\AA$\le\lambda\le$7.7$\,$\AA,~i.e.,~when the leading ion approaches the pore exit, reveals excellent consistency among $\rho(z,\lambda)$'s of the two larger systems. In contrast, the smallest system exhibits a qualitatively different behavior, with a leftward shift in the density peak. In other words, the partitioning of trailing chlorides into the pore is considerably curtailed in the smallest system. This can be attributed to the repulsive force exerted by the periodic replicates of the leading ion, as well as the attractive forces exerted by the periodic replicates of the positive image charges in the feed.  Such forces are significantly stronger in the smallest system due to the closer proximity of these periodic replicates to the trailing ion, meaningfully altering its spatial distribution within the pore. We refer to the arising artifacts as \emph{secondary finite-size effects} because they involve qualitative size-dependent changes in the distribution of non-traversing ions. Such secondary finite-size effects cannot be rectified by the \textsc{Icdm} model, which assumes that the spatial distribution of non-traversing ions does not exhibit a strong dependence on system size.

In principle, secondary finite-size effects can never be fully eliminated. However, observing the trailing charge distribution in these three systems reveals that these effects are only significant in the smallest system and negligible in the two larger systems. This  highlights the fundamental difference between primary and secondary finite-size effects. Primary finite-size effects are uniformly present and arise merely from charge polarization, regardless of system size. In contrast, secondary finite-size effects emerge abruptly when the simulation box is smaller than a specific threshold. In the case of the smallest system, we observe an abrupt hindrance in the partitioning of the trailing chloride  into the pore, which, in turn, slows down the traversal of the leading ion. Overlooking this qualitative change leads to the overestimation of the translocation barrier by the \textsc{Icdm} model. The model fails to properly quantify the knock-on effect of the trailing chloride on the leading ion, as it incorrectly assumes that the trailing chloride is further behind than it actually will be in the thermodynamic limit.

A key characteristics of all free energy profiles obtained in the $2\delta$ system is the presence of multiple peaks and valleys, which complicates the efficacy of the Arrhenius approach in estimating the passage time in the thermodynamic limit. Instead, the Markov state approach proposed in Section~\ref{section:Markov} is better suited for this system. While both approaches yield comparable passage times, the Markov state approach yields better statistical consistency between the two largest systems (that are devoid of secondary finite-size artifacts). There are, however, reasons to be skeptical of the performance of the Markov state model. Some of the corrected barriers are so close to $kT$ that the relevance of the Arrhenius approximation, which is used for estimating transition rates between successive Markov states, becomes questionable.

\subsubsection{Secondary translocation events}
\label{section:sodium}

\begin{figure*}[t]
	\centering
	\includegraphics[width=0.8\textwidth]{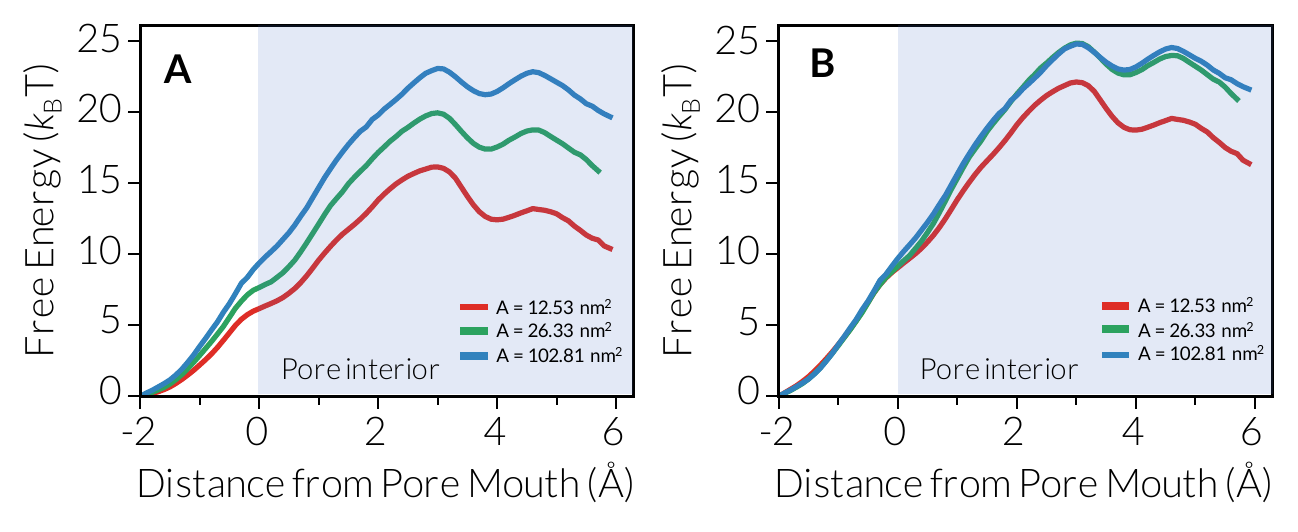}
	\caption{\label{fig:NaCorrections}(A) Uncorrected and (B) \textsc{Icdm}-corrected free energy profiles for the $F_{0,1} \rightarrow F_{1,1}$ transition in the $1\delta$ system. Shaded regions correspond to the pore interior.}
\end{figure*}

\noindent
In our earlier work,\cite{ShoemakerJCTC2022} we demonstrated that the \textsc{Icdm} model is capable of rectifying finite-size artifacts for secondary ion translocation events by considering the  $F_{0,1} \rightarrow F_{1,1}$ transition in the $1\delta$ system, which corresponds to the translocation of the first sodium ion after a chloride has already traversed the pore. This transition is  similar to the  $F_{0,0} \rightarrow F_{0,1}$ transition through the $2\delta$ system in the sense that in both cases, non-traversing ions can be present within the pore during the translocation process. Consequently, the emergence of secondary finite-size effects is, in principle, plausible in the $F_{0,1} \rightarrow F_{1,1}$ transition. However, we did not observe any such artifacts in the two system sizes that we considered in our earlier work.\cite{ShoemakerJCP2024} 

In order to examine this possibility, we probe secondary sodium transport in a considerably smaller system with a cross-sectional surface area of $12.53~\text{nm}^2$. The computed translocation times and barriers for all system sizes are provided in Table~\ref{tab:times-Na}, with the original and \textsc{Icdm}-corrected free energy profiles illustrated in Fig.~\ref{fig:NaCorrections}.  As expected, sodium transport is exponentially faster in the smallest system, since polarization-induced finite-size effects generally lead to an overestimation of translocation rate for co-ions.  As discussed in our prior work,\cite{ShoemakerJCP2024} this is due to the attractive force exerted on the leading sodium by periodic replicates of the lone chloride in the filtrate, which are stronger in smaller systems. Consequently, the \textsc{Icdm}-corrected barriers are all larger than the original barriers. We note the existence of strong finite-size effects for the system sizes considered here, with translocation rates and barriers varying by four orders of magnitude and  $\sim7\,kT$, respectively.

\begin{table*}
\caption{\label{tab:times-Na} Translocation timescales and free energy barriers of the $F_{0,1} \rightarrow F_{1,1}$ transition in the $1\delta$ system obtained from \textsc{jFfs} for different system sizes, along with the \textsc{Icdm}-corrected estimates.}
\begin{tabular}{crcccc}
\hline\hline
$A\,(\text{nm}^2)$ & $\tau_{f}\,(\text{s})$~~~ & $\beta \Delta \mathcal{F}_{f}$ &  $\beta \Delta \mathcal{F}_{\infty}$ & $\tau_{\infty,\text{Arrhenius}}\,(\text{s})$ & $\tau_{\infty,\textsc{Msm}}\,(\text{s})$ 
\\
\hline
~$12.53$ & $0.07\pm0.01$ & $16.15\pm`0.01$~ & $22.13\pm0.03$ & ~~~$27.42\pm2.84$ & ~$28.5\pm4.1$~~~\\
~$26.33$&  $6.72\pm0.83$ & $19.96\pm0.003$ & $24.73\pm0.01$ & ~$788.03\pm110$ & ~$871\pm108$\\
$102.81$ & $575.92\pm113$ & $23.09\pm0.02$~ & $24.78\pm0.07$ & $3127.68\pm872$ & $3263\pm642$\\
\hline 
\end{tabular}
\end{table*}

\begin{figure}[t]
	\centering
	\includegraphics[width=.5\textwidth]{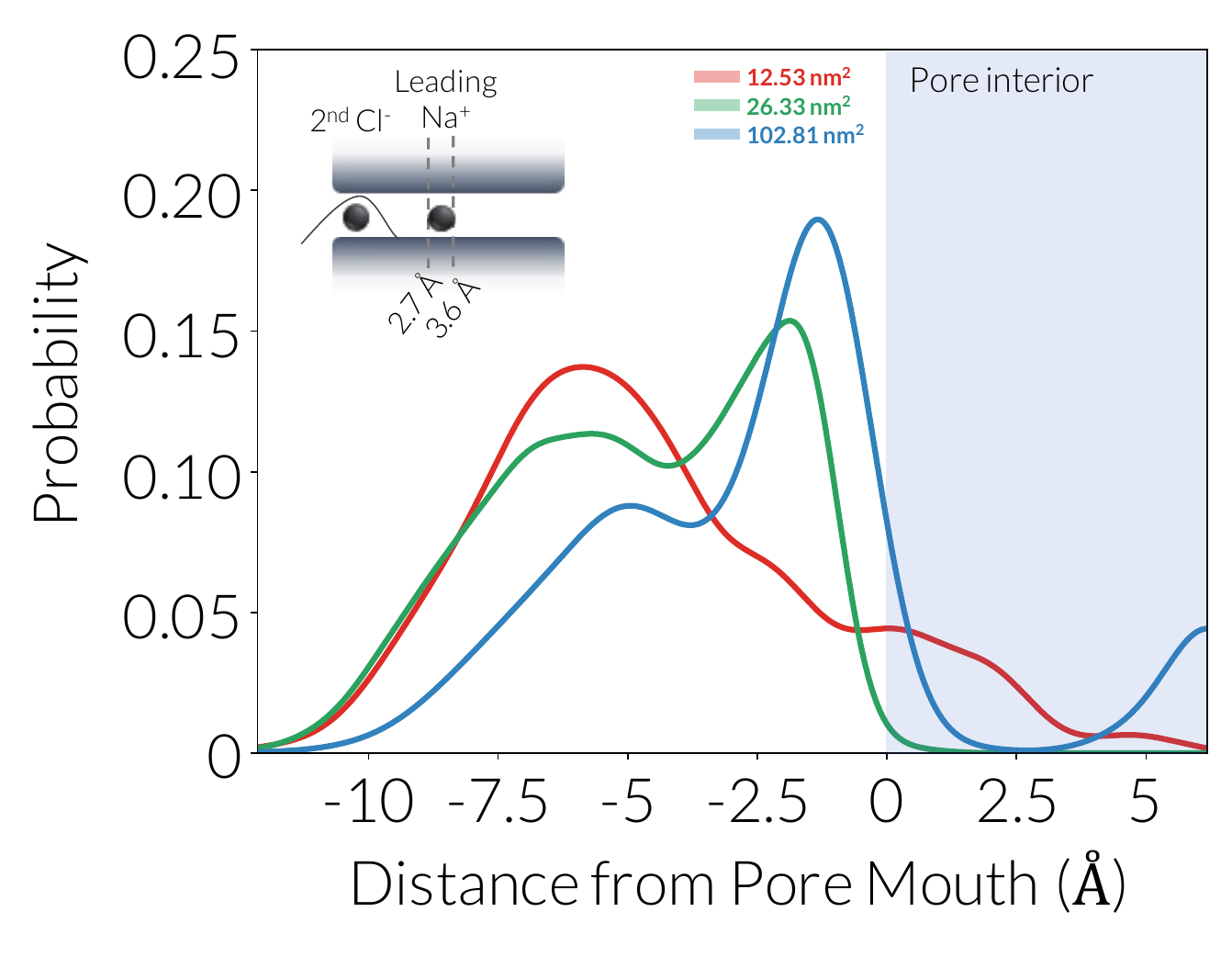}
	\caption{\label{fig:NaSys2ndClDist} Probability density of observing the second chloride at a specific $\lambda$ while the leading sodium is near the top of the translocation barrier for the $F_{0,1} \rightarrow F_{1,1}$ transition in the $1\delta$ system (2.7--3.6~\AA). The shaded region indicates the pore interior, and solid curves represent smoothing splines fitted to the underlying histograms. The inset illustrates the respective positions of the leading sodium and the second chloride.}
\end{figure}

Similar to chloride transport in the $2\delta$ system,  the \textsc{Icdm}-corrected free energy profile for the smallest system differs substantially from those of the two larger systems, which demonstrate remarkable consistency. More precisely, the \textsc{Icdm} model underestimates the translocation barrier in the thermodynamic limit by approximately $2.5\,kT$ in the smallest system,  suggesting the likely emergence of secondary finite-size effects. To reconcile these discrepancies and uncover the sources of such artifacts, we investigate the spatial distribution of secondary chloride  as the leading sodium  approaches the top of the barrier, as illustrated in Fig.~\ref{fig:NaSys2ndClDist}. Similar to the $2\delta$ system, the distribution changes meaningfully in the smallest system, becoming notably broader.  More specifically, the likelihood of locating a secondary chloride within the pore (i.e., for $z\ge0$) is substantially higher in the smallest system, although the chlorides in the pre-pore region are generally further from the pore mouth.  Conversely, in the two larger systems, the distribution is narrower with a pronounced chloride accumulation at the pore mouth.  This narrower distribution in the larger systems causes a deceleration of sodium transport due to the stronger attractive interaction between the trailing chloride and the traversing sodium. Ultimately, the qualitative difference in the trailing chloride distribution within the smallest system undermines the \textsc{Icdm} model's ability to accurately predict the free energy profile in the thermodynamic limit. As with chloride transport in the $2\delta$ system, these secondary finite-size effects are significant only in very small systems and can be neglected in sufficiently large systems.

An intriguing observation in the largest system concerns the locus of the tail of the trailing chloride distribution, which extends beyond the position of the leading sodium. This indicates that during the translocation process, a secondary chloride  can overtake the leading sodium  and proceed towards the pore exit. (We visually observe this in several of our \textsc{jFFS} crossing events.) This phenomenon suggests that the transport of the secondary chloride  may be considerably faster than that of the leading sodium in the thermodynamic limit. However, to confirm this finding, direct rate calculations for the $F_{0,1} \rightarrow F_{0,2}$ transition are necessary. Such calculations fall outside the scope of this work and will be addressed in future studies.

\subsection{How to avoid secondary finite-size effects?}
\label{section:heuristic}

\noindent
The complex nature and nontrivial origins of secondary finite-size effects make them exceptionally challenging to predict or rectify using first principles.  At a fundamental level, all polarization-induced finite-size artifacts stem from spurious interactions with the periodic replicates of ions within the system.  Primary artifacts, which can be fully accounted for and rectified using the \textsc{Icdm} model, occur due to interactions between the traversing ion and the periodic replicates of all non-traversing ions.  Conversely, secondary artifacts emanate from changes in the spatial distribution of non-traversing ions, spurred by their spurious interactions with the periodic replicates of the traversing ion. Such changes are not solely energetic and are influenced by various factors, such as hydration, interfacial effects, and ion-pore interactions. Moreover, secondary artifacts can  cascade into further artifacts that might be difficult to pinpoint or characterize. Formulating a universal theory to predict such changes in a system-agnostic manner is as challenging as analytically calculating partition functions in systems of interacting classical particles. Consequently, to accurately investigate the true physics of ion translocation, it is crucial to consider sufficiently large system sizes that are free from secondary artifacts, even if it comes with an increase in computational cost.

\begin{figure*}[t]
	\centering
	\includegraphics[width=1.0\textwidth]{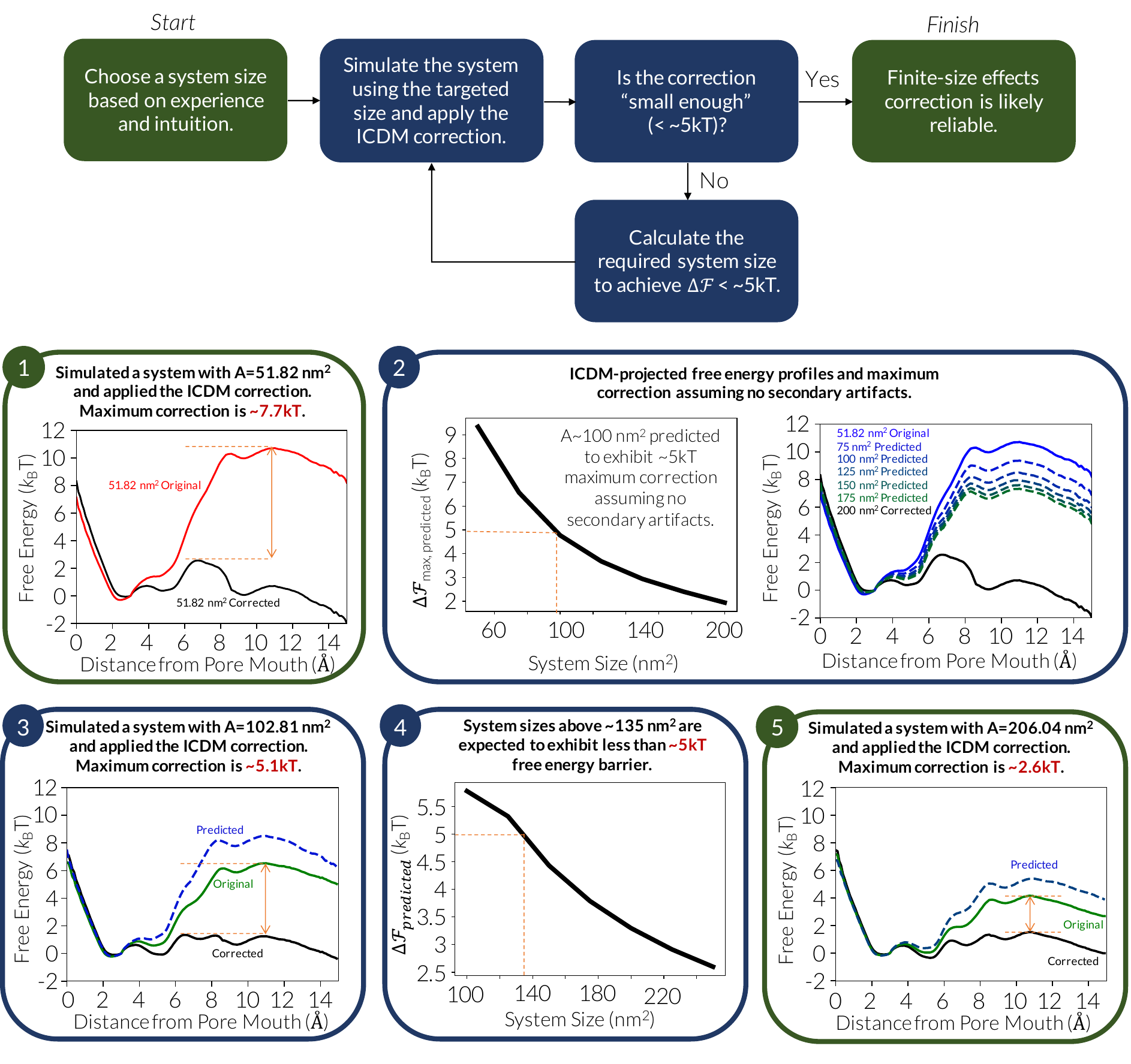}
	\caption{\label{fig:heuristics}A simple decision tree-based heuristic (top) to systematically gauge the extent of finite-size artifacts and the adequate system size threshold to avoid potential secondary effects. The panels numbered from 1 through 5 illustrates an actual implementation of the heuristic for the $2\delta$ system. Maximum correction to the free energy profile is defined as the difference in the magnitude of the largest barrier before and after the correction. The predicted profiles in panel 2 are produced using the \textsc{Icdm} model as discussed in the text. Predicted curves for step 4 are removed for conciseness. The iterations end when the maximum correction falls well below $5kT$.}
\end{figure*}

Our observation that secondary finite-size effects virtually disappear in sufficiently large systems suggests the existence of a system size threshold beyond which such artifacts will be negligible. However, since it is not feasible to predict the nature and extent of such secondary artifacts, such a threshold cannot be estimated rigorously, and will likely depend on the specifics of the system of interest. A vivid illustration of this is the differing sensitivity of the $F_{0,0}\rightarrow F_{0,1}$ transition to such artifacts in the $1\delta$ and $2\delta$ systems. Within the range of system sizes considered for the $1\delta$ system (12.53--102.81~nm$^2$), we observe no noticeable secondary artifacts. This is in contrast to the $2\delta$ system that exhibits strong secondary finite-size artifacts, which  only disappear  for system sizes approaching  $\approx100\,\text{nm}^2$. This higher susceptibility in the $2\delta$ system can be attributed to the tendency to facilitate  the entry of a second chloride into the pore, where the trailing chloride ion is less effectively screened and thus more impacted by the periodic replicates of the leading chloride. In contrast, the $1\delta$ pore does not typically accommodate a trailing chloride, explaining its robustness against significant secondary artifacts. Similarly, the stronger susceptibility of the $F_{0,1}\rightarrow F_{1,1}$ transition in the $1\delta$ system can be attributed to the existence of non-traversing ions within the pore.

Recognizing the inability to systematically and rigorously predict this threshold, we propose a heuristic workflow (outlined in Fig.~\ref{fig:heuristics}) to avoid secondary finite-size effects. Our approach is based on our intuitive expectation that secondary artifacts will be significantly less likely to occur if the magnitude of primary artifacts, as accounted for by the \textsc{Icdm} model, remains reasonably small. Although the precise definition of `reasonably small' remains ambiguous, our observations from this study suggest that a threshold of $5kT$ might be adequate, even though choosing a smaller threshold will undoubtedly lead to higher levels of confidence.

 The workflow begins with an initial rate and free energy calculation, where the system size is selected based on prior experience and physical intuition. The resulting free energy profile is subsequently corrected using the \textsc{Icdm} model. If the \textsc{Icdm} correction at the dominant free energy barrier is below the pre-determined threshold (e.g.,~$5kT$), the calculation can be considered `likely' free of secondary effects, and the \textsc{Icdm}-corrected barrier and rate can be utilized. Otherwise, the \textsc{Icdm} model is utilized to back-calculate the expected free energy profiles in successively larger systems. More precisely, for a system with box dimensions $L_{x}'$ and $L_{y}'$, the \textsc{Icdm}-projected free energy profile will be given by:
\begin{eqnarray}
&& \mathcal{F}_{f}^{\textsc{Icdm}}(z;L_{x}',L_{y,}') = \mathcal{F}_{f}^{\textsc{Md}}(z;L_x,L_y) \notag\\
&& - \Delta\mathcal{F}_{\text{corr}}(z,L_x,L_y)  + \Delta\mathcal{F}_{\text{corr}}(z,L_{x}',L_{y}'). \notag
\end{eqnarray} 
Here, $\mathcal{F}_{f}^{\textsc{Md}}(z;L_x,L_y)$ is the uncorrected free energy profile computed from the original simulation.
This formalism presumes that the implementation parameters of the \textsc{Icdm} model, including the membrane dielectric constant and the spatial distributions of non-traversing ions, do not change considerably upon altering the box dimensions. Therefore, the projected $\mathcal{F}_{\text{corr}}(z)$ can be  estimated by   simply increasing the lateral spacing between image charges and adjusting the surface charge densities of slab corrections. A larger system size can then be identified such that the maximum correction to the free energy barrier will fall below this threshold. A subsequent simulation can then be performed with the new system size, comparing its uncorrected translocation free energy profile to the predicted profile from the smaller system. If the profiles are sufficiently close, it indicates both calculations are likely devoid of secondary finite-size artifacts. Otherwise, this process will be repeated until the correction falls below the specified threshold. 

Figure~\ref{fig:heuristics} illustrates the application of this heuristic to primary chloride translocation in the $2\delta$ system. As can be seen in panel 3, there is a big discrepancy between the \textsc{Icdm}-projected free energy profile of the 102.81~nm$^2$ system, and the profile directly computed from \textsc{Md}, indicating the possibility of secondary finite-size artifacts. That discrepancy diminishes considerably in the subsequent iteration. We must, however, note that projected and actual free energy profiles will still differ slightly even in the absence of secondary finite-size effects, due to subtle sensitivity of \textsc{Icdm} parameters, such as the locus of the conductor surface and the membrane dielectric constant, to system size.

The practical heuristic outlined above offers a systematic framework for rectifying secondary finite-size artifacts. However, their existence--or likelihood thereof--can be inferred from certain indirect qualitative indications. One recurring theme among secondary artifacts discussed in this work is the intrusion of non-traversing ions into the membrane. Such intruding ions are not as effectively screened due to lower dielectric constant of membranes, increasing the likelihood of secondary artifacts. Relatedly, a strong charge imbalance between the two reservoirs will further exacerbate any secondary artifacts by facilitating or promoting such intrusion events. It is important to emphasize that ion intrusions do not constitute the only cause of such secondary artifacts, which might arise through other yet-to-be-determined mechanisms. Therefore, any drastic changes in the locus of the free energy barrier, or the topography of the free energy profile as a result of altering the system size \emph{might} be indicative of secondary finite-size effects. Moreover, if \textsc{Icdm}-corrected translocation timescales differ by more than an order of magnitude, secondary artifacts might be the culprits.

\subsection{Qualitative discrepancies and differing scaling of finite-size effects}

\noindent
In addition to providing accurate estimates of translocation times and barriers in a given system, \textsc{Md} simulations of hindered ion transport enable us to evaluate the qualitative impact of specific design variables on the kinetics and mechanisms of transport. The latter application is more valuable from a practical perspective, as the precise values of computed rates and barriers are often less relevant to realistic systems due to the intrinsic limitations of classical force fields.  Thus far, we have focused on the fact that polarization-induced finite-size effects render inaccurate estimates of rates and barriers. While such artifacts will be present irrespective of pore chemistry and geometry, their scaling with system size can depend heavily on such structural details. Consequently, ranking the qualitative behavior of systems that differ in pore chemistry and/or geometry but share the same cross-sectional surface area may yield misleading results that can be negated or altered with increasing the system size. This section demonstrates this concept through two illustrative examples.

First, we examine the effect of pore polarity on chloride transport by comparing the rates of $F_{0,0}\rightarrow F_{0,1}$ transition in the $1\delta$ and $2\delta$ systems. In the thermodynamic limit, the chloride passage time is approximately one nanosecond in the $2\delta$ system (Table~\ref{tab:times-2X}), which is markedly shorter than the 49 nanoseconds reported for the $1\delta$ pore in our previous work.\cite{ShoemakerJCP2024} This indicates that an increase in pore polarity enhances chloride transport by almost two orders of magnitude. (Note that the \textsc{Msm} model proposed here gives a slightly larger passage time in the $1\delta$ system, and therefore even a larger acceleration in the $2\delta$ system.) However, a qualitatively different pictures emerges upon comparing the transport kinetics in the 50~nm$^2$ systems: the passage time in the $2\delta$ pore, $4.29\pm0.23\,\mu$s, is an order of magnitude \emph{larger} than the $469\pm58$~ns passage time observed in the $1\delta$ system.\cite{ShoemakerJCP2024} It is only when the system size increases to approximately 100~nm$^2$ that the rank-order of passage times reverses, yielding $112\pm16$~ns and $91.93\pm4.28$~ns for the $1\delta$ and $2\delta$ pores, respectively. Even then, the difference between the two systems is considerably smaller than that observed in the thermodynamic limit. 

It is evident that polarization-induced corrections scale very differently with system size in these two pores, which differ solely in the partial charges of their functional groups while being otherwise identical. For instance, in the 102.81~nm$^2$ system, the  change in the height of the dominant barrier is approximately $0.5\,kT$ in the $1\delta$ system, but exceeds $5\,kT$ in the $2\delta$ system. This highlights the critical importance of meticulously applying finite-size corrections to both systems, as ignoring them could lead to erroneous conclusions, resulting in a mischaracterization of the impact of pore polarity on the kinetics and mechanism ion transport.

The second example involves the effect of pore arrangement on ion transport kinetics. The ionic flux through a multi-pore membrane comprised of $n_p$ uncorrelated pores will be the sum of fluxes through isolated pores, yielding a passage time $\tau_{\text{memb}}$ given by,
$$
\frac{1}{\tau_{\text{memb}}} = \sum_{i=1}^{n_p} \frac1{\tau_i},
$$
where $\tau_i$ is the passage time through the $i$-th pore. In a membrane comprised of $n_p$ identical pores, this will correspond to $\tau_{\text{memb}} = \tau_1/n_p$. If the pores are correlated, however, $\tau_{\text{memb}}$  will deviate from $\tau_1/n_p$, with the ratio $\mathcal{S}=n_p\tau_{\text{memb}}/\tau_1$ serving as a measure of pore-pore correlations. $\mathcal{S}>1$ will imply slowdown due to pore-pore correlations, while $\mathcal{S}<1$ will signify acceleration. In our previous work,\cite{ShoemakerACSNano2024} we examined the extent of such correlations for counter- and co-ion transport through $1\delta$ pores, and used the \textsc{Icdm} model to estimate $\mathcal{S}$ in the thermodynamic limit, observing slowdown in counter-ion and acceleration of co-ion transport. In Ref.~\citenum{ShoemakerACSNano2024}, all simulations were conducted in a system with a cross-sectional surface area of 102.81~nm$^2$, and although the computed slowdowns were modestly adjusted  upon applying the \textsc{Icdm} model, the simulated systems were large enough for the original estimates to be directionally accurate.

 In this work, we characterize the kinetics of $F_{0,0} \rightarrow F_{0,1}$ transition in the single- and double-pore $1\delta$ systems, but with a considerably smaller surface area of 32.60~nm$^2$. (The single-pore calculation was previously reported in our earlier works.\cite{ShoemakerJCTC2022,ShoemakerJCP2024}) For the larger 102.81~nm$^2$ systems, the computed passage times point to a slowdown by a factor of $\mathcal{S}=2.04\pm0.29$. Conversely, the smaller systems exhibit the opposite trend, where the per-pore ion transport is substantially faster in the double-pore system. The passage time is $584\pm27$~ns in the double-pore system, which is almost six times faster than $3.38\pm0.17\,\mu$s in the single-pore system.  Moreover, the translocation barrier in the double-pore system, $(8.80\pm0.02)\,kT$, is smaller than the single-pore system $(9.41\pm0.01)\,kT$. Similar to the first example, finite-size artifacts yield an inaccurate picture of pore-pore correlations, underscoring the importance of making comparisons with simulations conducted in sufficiently large systems.

\section{Conclusions}
\label{section:conclusions}

\noindent
In this work, we uncover a new class of polarization-induced finite-size effects in molecular simulations of hindered ion transport. These  artifacts have the same origin as the  finite-size effects that we had discovered in our earlier works,\cite{ShoemakerJCTC2022, ShoemakerJCP2024}  and result from spurious long-range electrostatic interactions among ions and their lateral periodic replicates. The original artifacts-- which we call primary-- only influence the force exerted on the traversing ion. When such primary artifacts get stronger upon decreasing the system size, they can alter the spatial distribution of non-traversing ions in a manner that deviates from the thermodynamic limit. This could, for instance, happen if the translocation process involves the simultaneous intrusion of multiple ions into the pore. Then, the distribution of non-traversing ions will change (e.g.,~due to periodic replicates of the traversing ion) and that will in turn affect the force exerted on the leading ion. These artifacts-- which we call secondary-- cannot be rectified by the \textsc{Icdm} model, which operates under the assumption that non-traversing ion distribution within and near the pore does not change considerably compared to the thermodynamic limit. As such, the \textsc{Icdm}-corrected passage times and translocation barriers will exhibit considerable systematic bias.

Unlike primary artifacts that persist for all system sizes, secondary artifacts tend to vanish for systems that are sufficiently large. Considering their complex origins, and the difficulties of quantitatively mitigating them in a system-agnostic manner, it is important to avoid them by simulating systems that are large enough to be devoid of secondary artifacts. However, it is nontrivial to determine such a threshold, which is often system-specific, and is determined by an interplay between the energetics of ion intrusion and the magnitude of spurious interactions. We therefore propose a simple heuristic that assures that the magnitude of the maximum \textsc{Icdm} correction does not exceed a pre-determined threshold. Systems wherein the \textsc{Icdm} correction exceeds this threshold are considerably likelier to be susceptible to secondary artifacts. 

In addition to discovering secondary finite-size effects, we propose a new dynamical approach for estimating translocation times in the thermodynamic limit. This approach, which is a generalization of our original Arrhenius-based approach, treats ion translocation as a Markov process wherein the valleys of the free energy profile constitute the Markov states, with the target basin as an absorbing state. By assuming the validity of the Arrhenius picture among every two successive basins, transition matrices are constructed for both the original and \textsc{Icdm}-corrected free energy profiles, and their spectral properties are compared to estimate the degree of slowdown or acceleration.

We also demonstrate  through multiple examples that finite-size effects could exhibit vastly different scalings with system size if the pore chemistry and/or geometry is altered. Therefore, rank-ordering  the qualitative behavior of systems that differ in terms of pore chemistry or arrangement but share the same cross-sectional surface area might yield qualitatively inaccurate and misleading conclusions.  This underscores the importance of meticulously identifying possible secondary artifacts in ion transport simulations, and applying the \textsc{Icdm} model to correct for well-defined primary artifacts. 

One of the main reasons that exacerbates primary and secondary artifacts in the types of membranes considered here-- and in many other studies-- is the relatively low dielectric constants of such membranes. Real synthetic and biological membranes often possess much higher dielectric constants, and can thus screen spurious long-range interactions responsible for polarization-induced finite-size artifacts. One practical solution could involve representing membrane regions as polarizable media with desired dielectric constants, using methods such as Drude oscillators.\cite{LamoureuxJCP2003} Such approaches are often more expensive computationally, but might offer added certainty about the robustness of one's conclusions to finite-size effects. Moreover, since finite-size effects often slow down counter-ion transport, using Drude oscillators might lead to an overall decrease in computational cost (due to orders of magnitude faster translocation timescales).

\begin{acknowledgments}

\noindent
This work was supported as part of the Center for Enhanced Nanofluidic Transport (\textsc{Cent}$^2$), an Energy Frontier Research Center funded by the U.S. Department of Energy, Office of Science, Basic Energy Sciences under Award \#DE-SC0019112. A.H.-A. gratefully acknowledges the support of the National Science Foundation Grants \textsc{Cbet}-1751971 (\textsc{Career} Award) and \textsc{Cbet}-2024473. B.S. acknowledges the support of the Goodyear Tire \& Rubber Fellowship. These calculations were performed on the Yale Center for Research Computing. This work used the Extreme Science and Engineering Discovery Environment (\textsc{Xsede}), which is supported by National Science Foundation grant no. \textsc{Aci}-1548562. This work used Stampede through allocation \textsc{Chm}240063 from the Advanced Cyberinfrastructure Coordination Ecosystem: Services \& Support (\textsc{Access}) program, which is supported by \textsc{Nsf} Grants \#2138259, \#2138286, \#2138307, \#2137603, and \#2138296.
\end{acknowledgments}

\bibliographystyle{apsrev} 
\bibliography{References-all}

\end{document}